\documentclass[preprint,prc,aps,nofootinbib,superscriptaddress]{revtex4}
\usepackage{graphicx}
\usepackage{epsfig}
\usepackage{bm}
\usepackage{amssymb}

\newcommand{\lag}{\mathcal L}

\newcommand{\be}{\begin{eqnarray}}
\newcommand{\ee}{\end{eqnarray}}
\begin{document}

\title{Medium effects of magnetic moments of baryons on neutron stars under
strong magnetic fields}

\author{C. Y. Ryu} \email{cyryu@skku.edu}
\affiliation{Department of Physics,
Soongsil University, Seoul 156-743, Korea}

\author{K. S. Kim} \email{kyungsik@kau.ac.kr}
\affiliation{School of Liberal Arts and Science, Korea Aerospace
University, Koyang 412-791, Korea}

\author{Myung-Ki Cheoun} \thanks{Corresponding Author} \email{cheoun@ssu.ac.kr}
\affiliation{Department of Physics,
Soongsil University, Seoul 156-743, Korea}


\begin{abstract}
We investigate medium effects due to density-dependent magnetic
moments of baryons on neutron stars under strong magnetic fields.
If we allow the variation of anomalous magnetic moments (AMMs) of
baryons in dense matter under strong magnetic fields, AMMs of
nucleons are enhanced to be larger than those of hyperons. The
enhancement naturally affects the chemical potentials of baryons
to be large and leads to the increase of a proton fraction.
Consequently, it causes the suppression of hyperons, resulting in
the stiffness of the equation of state. Under the presumed strong
magnetic fields, we evaluate relevant particles' population, the
equation of state and the maximum masses of neutron stars by
including density-dependent AMMs and compare them with those
obtained from AMMs in free space.
\end{abstract}
\maketitle

\section{Introduction}
Recently, strong magnetic fields were observed at the surface of
soft gamma ray repeaters, called magnetar. The magnitude of the
fields was estimated as an order of $10^{14} - 10^{15}$ G
\cite{cardall2001}. In the interior of neutron stars, according to
the scalar virial theorem, the magnetic field strength could
be about $10^{18}$ G. Such strong magnetic fields may affect
the structure of a neutron star such as the populations of particles, the equation of state (EoS)
and mass-radius relations. Many studies for neutron stars with strong
magnetic fields have been reported by several papers, which
included the electromagnetic interaction, the Landau quantization
of charged particles, and anomalous magnetic moments (AMMs) of
baryons
\cite{band1997,brod2000,su2001,dey2002,shen2006,shen2009,panda2009}.
But roles of relevant particles' AMMs in a strong magnetic field
are still uncertain because properties of the AMMs in nuclear
matter are not fully scrutinized yet.

On the other hand, medium effects of the electromagnetic (EM) form
factors for nucleons have been mainly investigated on the electron
scattering in both experimental \cite{emc83,mulders90,jlab07} and
theoretical aspects \cite{kc03,kw03}. From their results, one may
expect the swollen effect of the EM form factors by about 20
$\sim$ 40 \%. In particular, various possible variations of the
AMMs of baryons in nuclear matter have been studied extensively by
many different theoretical models
\cite{meisner,cheon92,frank96,lu99,yak03,smith04,horikawa05,ccts2008}.
However, there are still remained some ambiguities on the density
dependence of the AMMs stemming from the model dependence of
baryons in nuclear matter. Furthermore, experimental data also
show large error bars. For example, the AMM of $\Lambda$ hyperon
in $^{7}_{\Lambda}$Li nucleus recently measured at BNL
\cite{tamura07} still showed large error bars. Further experiments
are expected to deduce more clearly the AMM properties in nuclear
matter.

Authors in Ref. \cite{ccts2008} studied the medium dependence of the AMMs of
baryons in symmetric nuclear matter by using both different models, the
quark-meson coupling (QMC) \cite{guichon88} and the modified
quark-meson coupling (MQMC) models \cite{mqmc96}. In the QMC
model, the density dependence of the AMMs of baryons is very
small, while the AMM values of a proton and a $\Lambda$ hyperon in
the MQMC model are enhanced by about $25 \%$ and about $10 \%$,
respectively, at saturation density. Such large enhancements in
the MQMC model are quite feasible because the AMM of a baryon
generally depends strongly on the bag radius.

In the sense, the MQMC model could effectively take the swollen
effect of nucleons into account, by increasing the bag radius about
$20 \%$ at saturation density. But in the QMC model, the bag
radius is rarely changed to make the change of AMMs very small.
Therefore, the MQMC model can provide us with a theoretical framework
to discuss medium effects of the AMMs.

In this work, under the assumption that the AMM values of baryons
may considerably depend on medium, we apply the effects to a neutron star.
The calculation of the AMMs of baryons in medium is done by considering only
SU(6) quark wave functions obtained by using the MQMC model.
Further possible consequences of the effects under strong
magnetic fields are also discussed
from observational quantities on neutron stars .

Since the quantum hadrodynamics (QHD), which is a systematically
developed model for finite nuclei and nuclear matter, provides us with
results very similar to those by the MQMC model for the structure of a neutron star, we
employ the QHD model for a neutron star under strong magnetic
fields by including the electromagnetic potential, the Landau
quantization of charged particles, and the AMM values of baryons
\cite{brod2000,dey2002,shen2009,panda2009}. But, to extract the
density dependence of the AMMs, we adopt the MQMC model because
the model can be easily applied to describe the AMMs in nuclear
matter rather than the QHD model and generate successfully the AMM
values of baryon octets in nuclear matter.

This paper is organized as follows. In Sec. II, the QHD model for
dense matter under a strong magnetic field is briefly introduced
by focusing on the roles of the AMM in the magnetic field. Results
and discussions are presented in Sec. III. Summary and conclusions
are given at Sec. IV.

\section{Theory }

The lagrangian density of the QHD model for dense matter in the presence of strong
magnetic fields, which is introduced by the vector potential $A^{\mu}$ due
to magnetic fields, can be represented in terms of octet
baryons, leptons, and five meson fields as follows
\be \lag &=& \sum_b \bar \psi_b \Big [ i \gamma_\mu \partial^\mu
 - q_b \gamma_\mu A^\mu - M_b^*(\sigma, \sigma^*) - g_{\omega b} \gamma_\mu \omega^\mu
 - g_{\phi b} \gamma_\mu \phi^\mu  \nonumber \\
&& - g_{\rho b} \gamma_\mu \vec \tau \cdot \rho^\mu
 - \frac 12 \kappa_b \sigma_{\mu \nu} F^{\mu \nu}
\Big ] \psi_b  
 + \sum_l \bar \psi_l \left [  i \gamma_\mu \partial^\mu
 - q_l \gamma_\mu A^\mu - m_l \right ] \psi_l  \nonumber \\
&& + \frac 12 \partial_\mu \sigma \partial^\mu \sigma - \frac 12 m_\sigma^2 \sigma^2 - U(\sigma)
+ \frac 12 \partial_\mu \sigma^* \partial^\mu \sigma^* - \frac 12 m_{\sigma^*}^2 {\sigma^*}^2
- \frac 14 W_{\mu \nu}W^{\mu \nu} \nonumber \\
&& + \frac 12 m_\omega^2 w_\mu w^\mu
- \frac 14 \Phi_{\mu \nu} \Phi^{\mu \nu} + \frac 12 m_\phi^2 \phi_\mu \phi^\mu
- \frac 14 R_{i \mu \nu} R_i^{\mu \nu} + \frac 12 m_\rho^2 \rho_\mu \rho^\mu
- \frac 14 F_{\mu \nu} F^{\mu \nu},
\ee
where $b$ and $l$ denote the octet baryons and the leptons ($e^-$
and $\mu^-$), respectively. The effective mass of a baryon,
$M_b^*$, is simply given by $M_b^* = M_b - g_{\sigma b}\sigma -
g_{\sigma^* b} \sigma^*$, where $M_b$ is the free mass of a baryon
in vacuum. The $\sigma$, $\omega$ and $\rho$ meson fields describe
interactions of nucleon-nucleon ($N-N$) and nucleon-hyperon
($N-Y$). Interaction of $Y-Y$ is mediated by $\sigma^*$ and $\phi$
meson fields. $U(\sigma)$ is the self interaction of the $\sigma$
field given by $U(\sigma) = \frac 13 g_2 \sigma^3 + \frac 14 g_3
\sigma^4$. $W_{\mu\nu}$, $R_{i \mu \nu}$, $\Phi_{\mu\nu}$, and
$F_{\mu\nu}$ represent the field tensors of $\omega$, $\rho$,
$\phi$ and photon fields, respectively. The AMMs of baryons
interact with an external magnetic field in the form of $\kappa_b
\mu_N \sigma_{\mu\nu} F^{\mu\nu}$ where $\sigma_{\mu \nu} = \frac
i2 [\gamma_\mu, \gamma_\nu]$ and $\kappa_b$ is the strength of AMM
of a baryon, i.e. $\kappa_p = 1.7928 \mu_N$ for a proton in a vaccum where
$\mu_N$ is the nucleon magneton defined as $\mu_N = e / 2m_p$. Since the medium dependence of the AMM on the density is considered, the ratios of $\kappa_b$ of baryon octet are taken from the results of the MQMC model in Ref. \cite{ccts2008}.
In the MQMC model, baryons are treated as MIT bags
and $\kappa_b$ is calculated from SU(6) quark wave functions and the bag radius depending on medium. For $\mu_N = e / 2m_p$, $\mu_N$ is always defined with
the mass of a proton in free space, $m_p$. Thus in this work, $\kappa_b$ depends on the density but $\mu_N$ does not.

The Dirac equations of octet baryons and leptons in the mean field
approximation are given by \be \Big [ i \gamma_\mu
\partial^\mu &-& q_b \gamma_\mu A^\mu - M_b^*(\sigma, \sigma^*) -
g_{\omega b} \gamma^0 \omega_0
 - g_{\phi b} \gamma^0 \phi_0  \nonumber \\
&-& g_{\rho b} \gamma^0 \tau_3 \rho_{30}
 - \frac 12 \kappa_b \sigma_{\mu \nu} F^{\mu \nu} \Big ] \psi_b = 0, \ee
\be
(i \gamma_\mu \partial^\mu
 - q_l \gamma_\mu A^\mu - m_l ) \psi_l = 0,
\ee where $A_\mu = (0, ~0, ~Bx, ~0)$ refers to the constant
magnetic field $B$, which is assumed as along the $z$-axis. The
energy spectra of baryons and leptons are given by
\be E_b^C &=& \sqrt{k_z^2 + \left ( \sqrt{{M_b^*}^2 + 2 \nu |q_b|
B} - s \kappa_b B \right )^2}
+ g_{\omega b}\omega_0 + g_{\phi b} \phi_0 + g_{\rho b} I_3^b \rho_{30}, \nonumber \\
E_b^N &=& \sqrt{k_z^2 + \left ( \sqrt{{M_b^*}^2 + k_x^2 + k_y^2} - s \kappa_b B \right )^2}
+ g_{\omega b}\omega_0 + g_{\phi b} \phi_0 + g_{\rho b} I_3^b \rho_{30}, \nonumber \\
E_l &=& \sqrt{k_z^2 + m_l^2 + 2 \nu |q_l| B},
\ee
where $E_b^C$ and $E_b^N$ represent energies of a charged baryon
and a neutral baryon, respectively. The Landau quantization of a
charged particle due to magnetic fields is denoted as $\nu = n
+ 1/2 - sgn(q) s/2 = 0, 1, 2 \cdots~$ with electric charge $q$ and
spin up (down) $s=1(-1)$. Equations of meson fields are given by
\be m_\sigma^2 \sigma + \frac{\partial U(\sigma)}{\partial
\sigma}&=& g_{\sigma b}
\sum_b \rho_s^b ,\nonumber \\
m_{\sigma^*}^2 \sigma^* &=& g_{\sigma^* b} \sum_b \rho_s^b, \nonumber \\
m_\omega^2 \omega_0 &=& g_{\omega b} \sum_b \rho_v^b, \nonumber \\
m_\phi^2 \phi_0 &=& g_{\phi b} \sum_b \rho_v^b, \nonumber \\
m_\rho^2 \rho_{30} &=& g_{\rho b} \sum_b I_3^b \rho_v^b,
\label{eq:mesons} \ee
where $\rho_s$ and $\rho_v$ are the scalar and the vector
densities under magnetic fields, respectively. Detail
expressions for these quantities are given in Ref.
\cite{brod2000,shen2009}. The chemical potentials of baryons and
leptons are, respectively, given by \be \mu_b &=& E_f^b +
g_{\omega b}\omega_0 + g_{\phi b} \phi_0 + g_{\rho b} I_z^b
\rho_{30}, \label{eq:che-b}
\\
\mu_l &=& \sqrt{k_f^2 + m_l^2 + 2 \nu |q_l| B}, \label{eq:che-l}
\ee
where $E_f^b$ is the Fermi energy of a baryon and $k_f$ is the
Fermi momentum of a lepton. For charged particles, the $E_f^b$ is
written as
\be {E_f^b}^2 = {k_f^b}^2 + (\sqrt{{m_b^*}^2 + 2 \nu |q_b| B} - s
\kappa_b B)^2,
\ee
where $k_f^b$ is the Fermi momentum of a baryon. Since the Landau
quantization does not appear for neutral baryons, the Fermi energy
is simply given by
\be {E_f^b}^2 = {k_f^b}^2 + ({m_b^*} - s \kappa_b B)^2.
\ee

We exploit three constraints for calculating properties of a
neutron star: baryon number conservation, charge neutrality, and
chemical equilibrium. The meson field equations in Eq.
(\ref{eq:mesons}) are solved with the chemical potentials of
baryons and leptons under the above three constraints. Total
energy density is given by $\varepsilon_{tot} = \varepsilon_m +
\varepsilon_f$, where the energy density for matter fields is
given by
\be \varepsilon_m &=& \sum_b \varepsilon_b + \sum_l \varepsilon_l
+ \frac 12 m_\sigma^2 \sigma^2 + \frac 12 m_{\sigma^*}^2
{\sigma^*}^2 + \frac 12 m_\omega^2 \omega^2 + \frac 12 m_\phi^2
\phi^2 + \frac 12 m_\rho^2 \rho^2 + U(\sigma),
\ee
and the energy density due to the magnetic field is given by
$\varepsilon_f = B^2 / 2$. The total pressure can also be written
as \be P_{tot} = P_m + \frac 12 B^2, \ee where the pressure due to
matter fields is obtained from $P_m = \sum_i \mu_i \rho_v^i -
\varepsilon_m$. The relation between mass and radius for a static
and spherical symmetric neutron star is generated by calculating
the Tolman-Oppenheimer-Volkoff (TOV) equations with the equation
of state (EoS) above.

\section{Results and discussion}

We use the parameter set in Ref. \cite{RHHK} for the coupling
constants, $g_{\sigma N}$, $g_{\omega N}$ and $g_{\rho N}$, where
$N$ denotes the nucleon. For the coupling constants of hyperons in
nuclear medium, $g_{\omega Y}$ is determined by the quark counting
rule, and $g_{\sigma Y}$ is fitted to reproduce the potential of
each hyperon at saturation density, whose strengths are given by
$U_\Lambda = -30$ MeV, $U_\Sigma = 30$ MeV and $U_\Xi = -15$ MeV.
For density-dependent AMM values of baryons, we use the values
obtained from our previous calculation done by the MQMC model
\cite{ccts2008}. Since the magnetic fields may also depend on
density, we take density-dependent magnetic fields used in
Refs. \cite{band1997,panda2009}
\be B \left ( \rho / \rho_0 \right ) = B^{surf} + B_0 \left [ 1 -
\exp \{-\beta \left ( \rho / \rho_0 \right )^\gamma \} \right ],
\label{eq-B} \ee
where $B^{surf}$ is the magnetic field at the surface of a neutron
star, which is taken as $10^{15}$ G from observations and $B_0$
represents the magnetic field saturated at high densities.

In the present work, we use two different sets, slow ($\beta =
0.05$ and $\gamma = 2$) and fast ($\beta = 0.02$ and $\gamma = 3$)
varying magnetic fields. Since the magnetic field is usually
written in a unit of the critical field for the electron $B_e^c =
4.414 \times 10^{13}$ G, the $B$ and the $B_0$ in Eq. (\ref{eq-B})
can be written as $B^* = B / B_e^c$ and $B_0^* = B_0 /B_e^c$.
Here, we regard the $B_0^*$ as a free parameter and investigate
medium effects of AMMs in a neutron star for three different
magnetic fields given by $B_0^* = 1 \times 10^5$, $2 \times 10^5$,
and $3 \times 10^5$.

\subsection{Medium effects on the populations of particles}

\begin{figure}
\centering
\includegraphics[width=7.8cm]{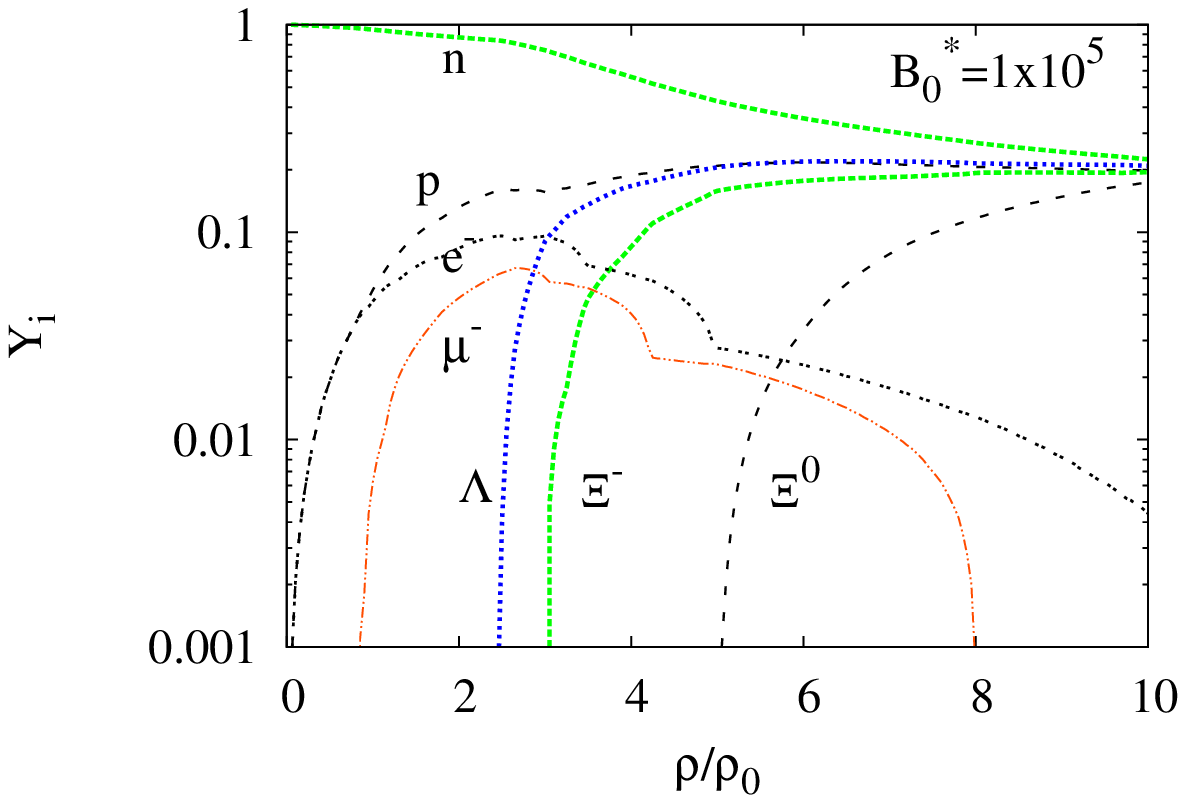}
\includegraphics[width=7.8cm]{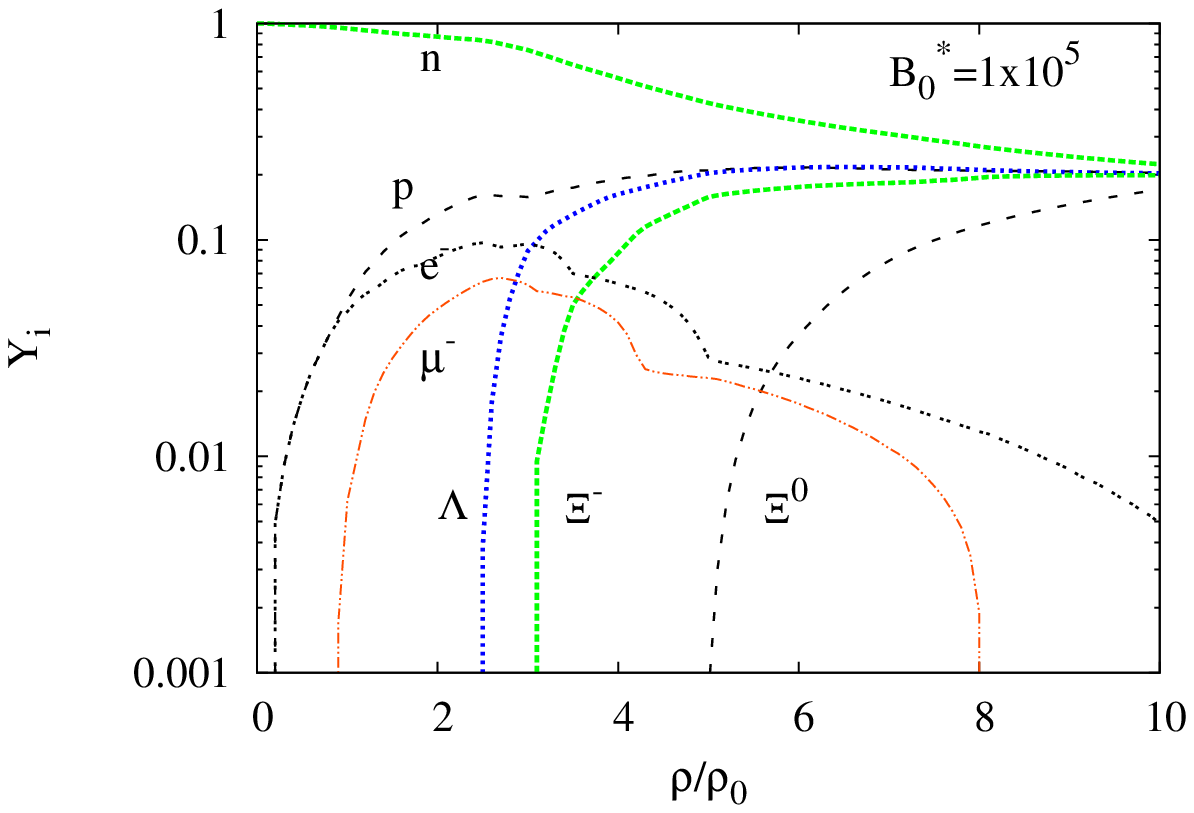}  \\
\includegraphics[width=7.8cm]{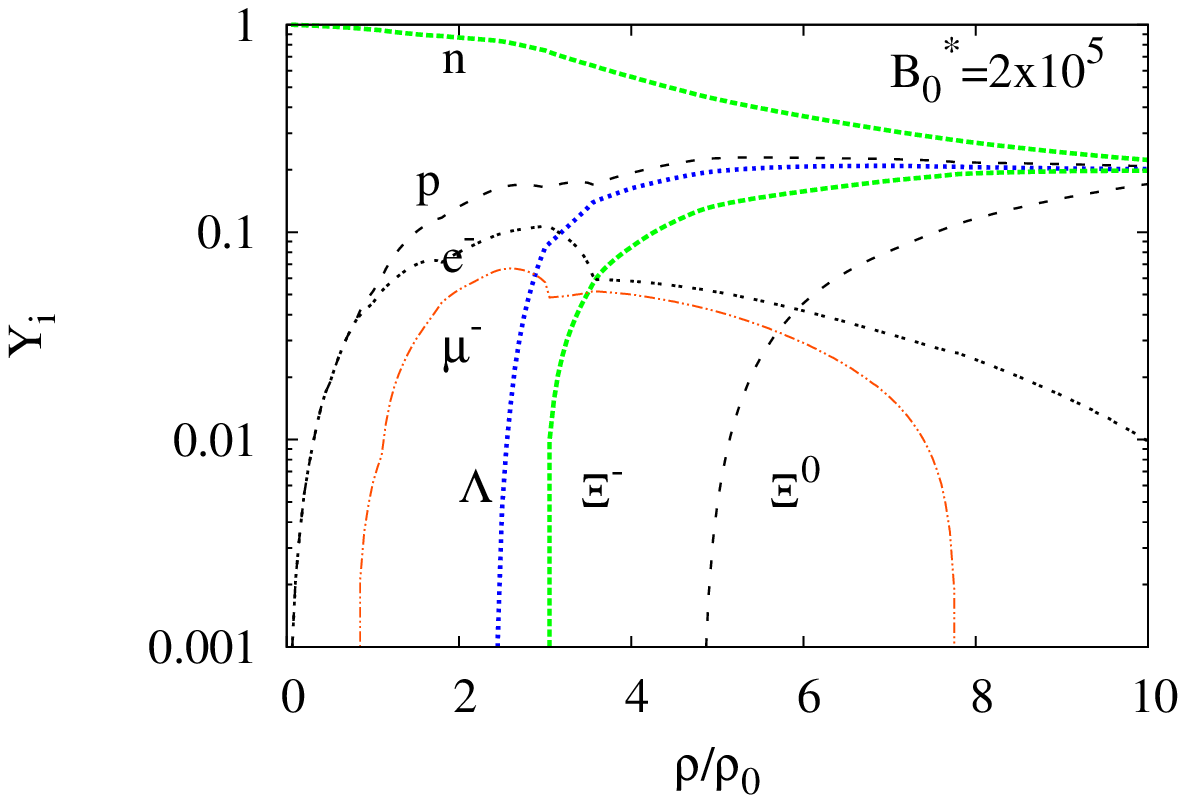}
\includegraphics[width=7.8cm]{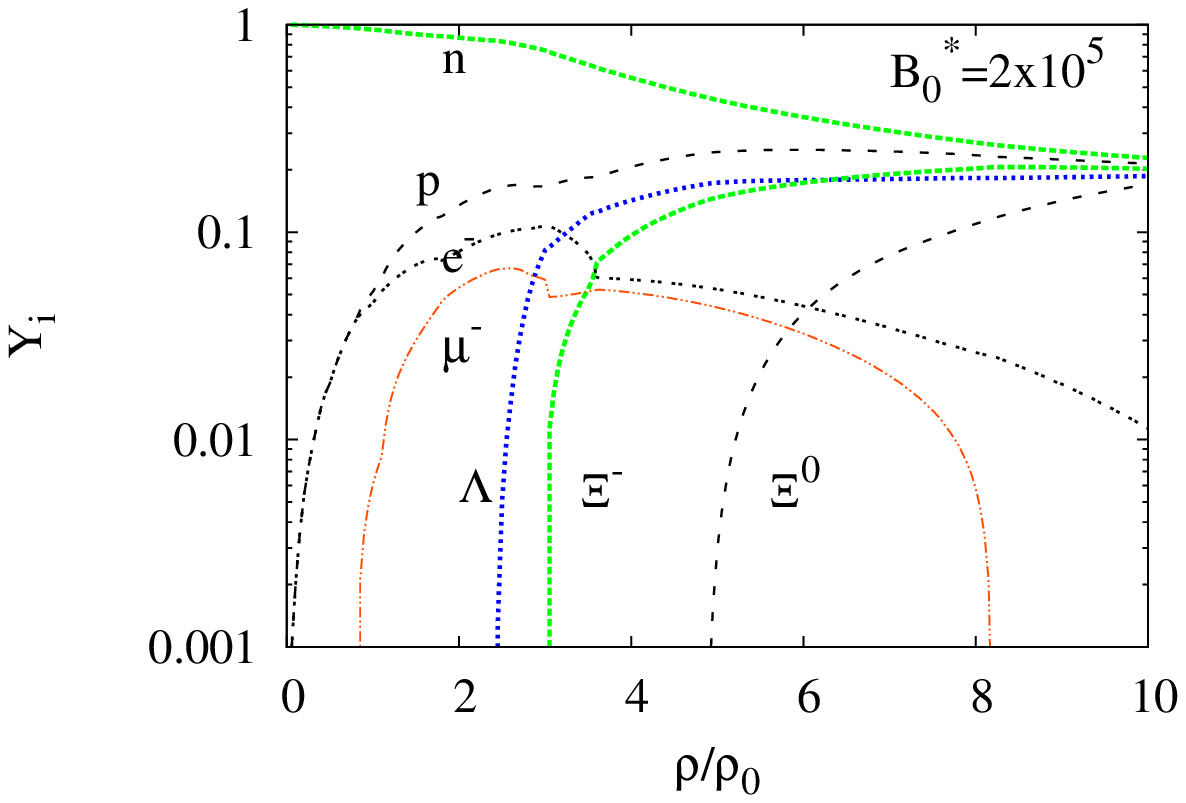}  \\
\includegraphics[width=7.8cm]{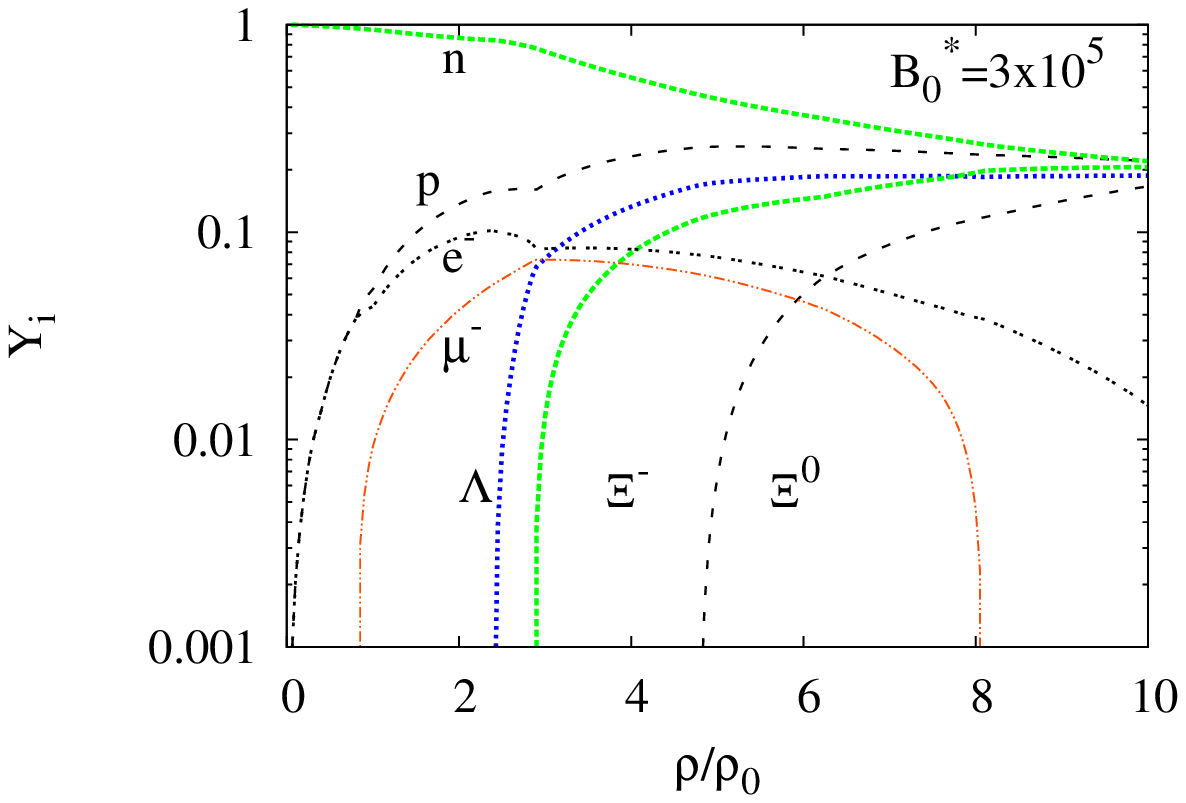}
\includegraphics[width=7.8cm]{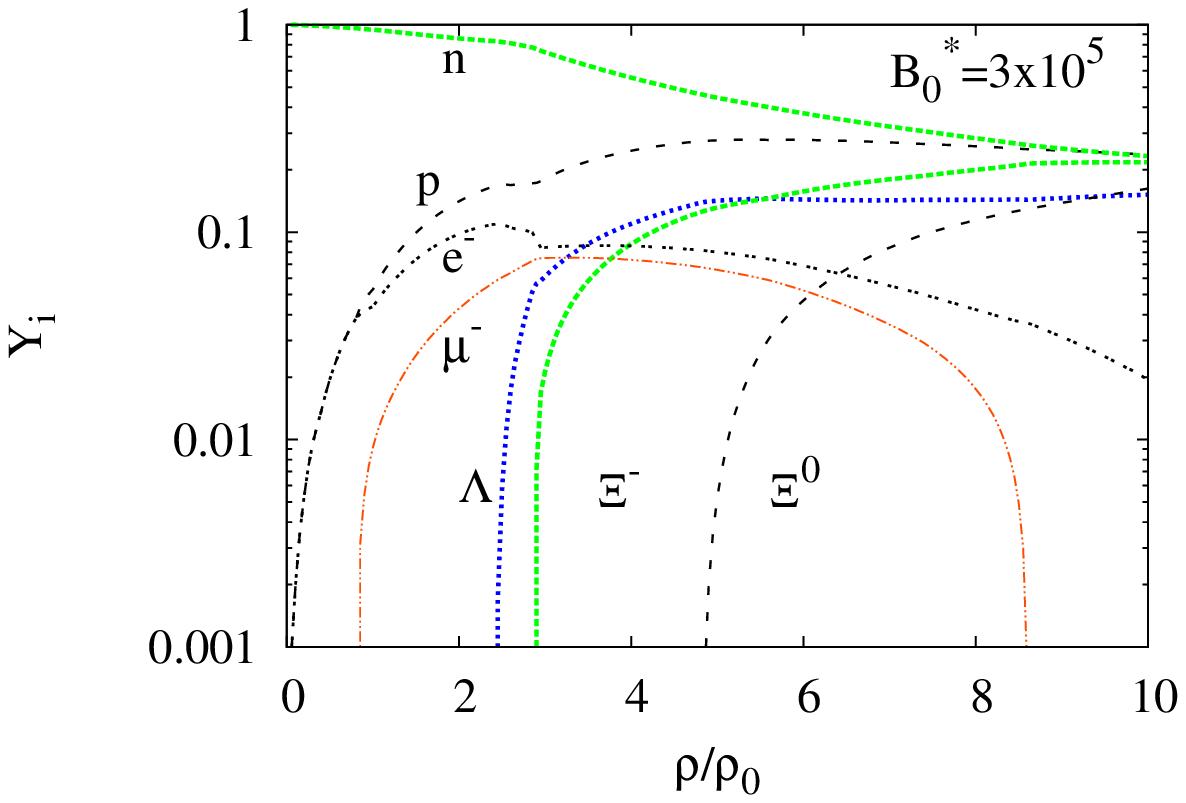}
\caption{(Color online) Populations of particles in a neutron star for the slow
varying magnetic field ($\beta = 0.05$ and $\gamma = 2$). Left
panels denote results for constant AMMs in free space and right
panels are for density-dependent AMMs obtained from the MQMC
model. For more direct comparison, all results for p and $\Xi^-$
are summarized in the left hand side (LHS) of Fig.3.}
\label{fig:popul-s}
\end{figure}
\begin{figure}
\centering
\includegraphics[width=7.8cm]{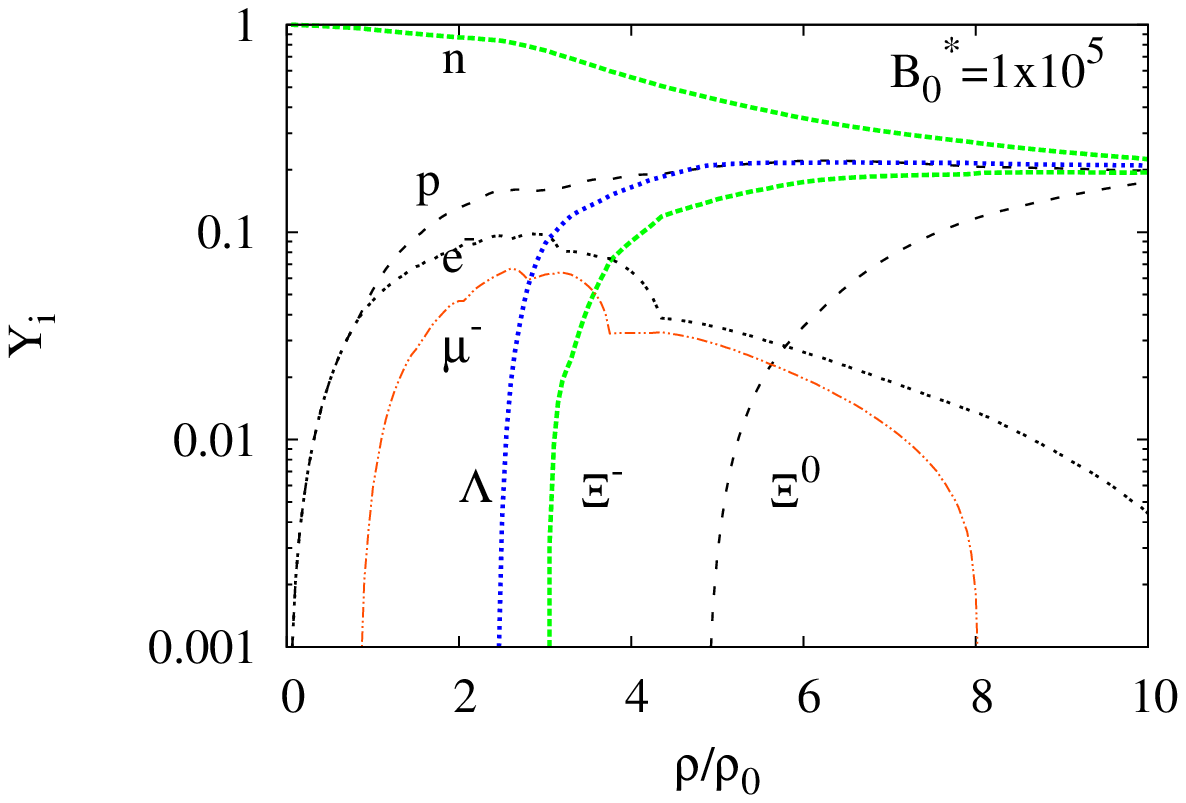}
\includegraphics[width=7.8cm]{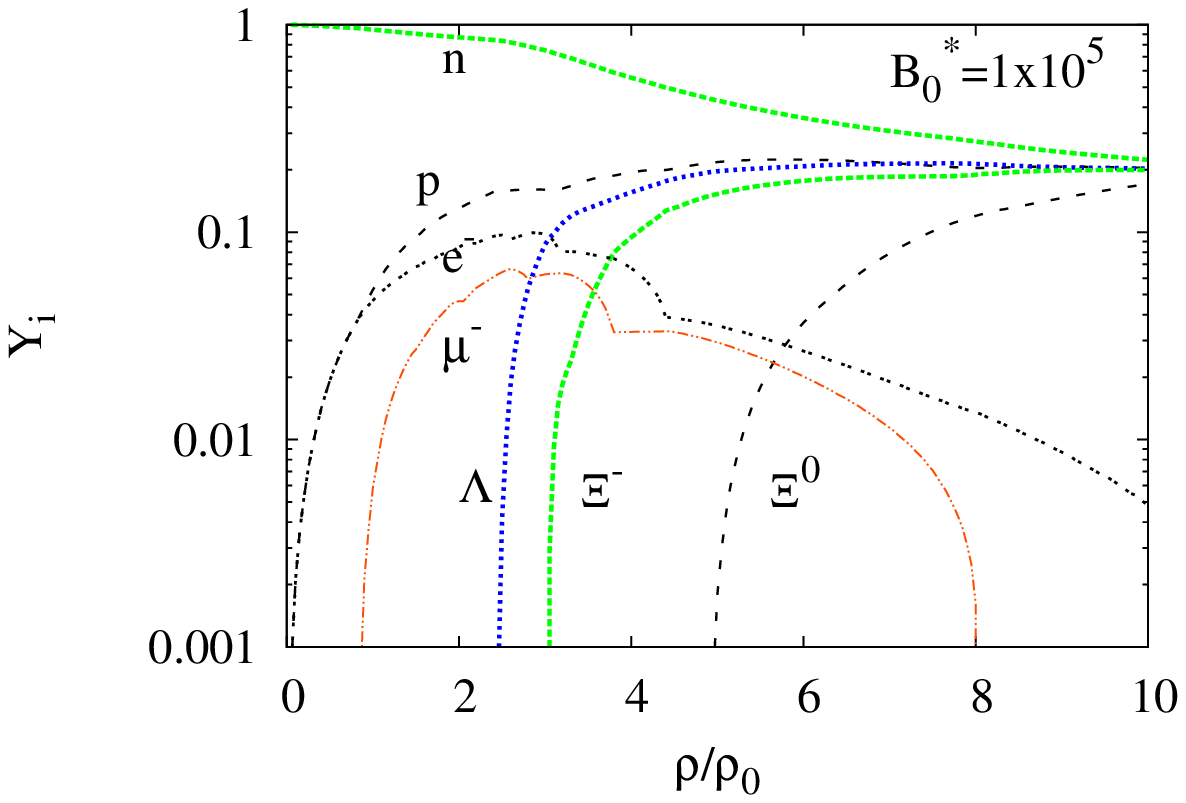}  \\
\includegraphics[width=7.8cm]{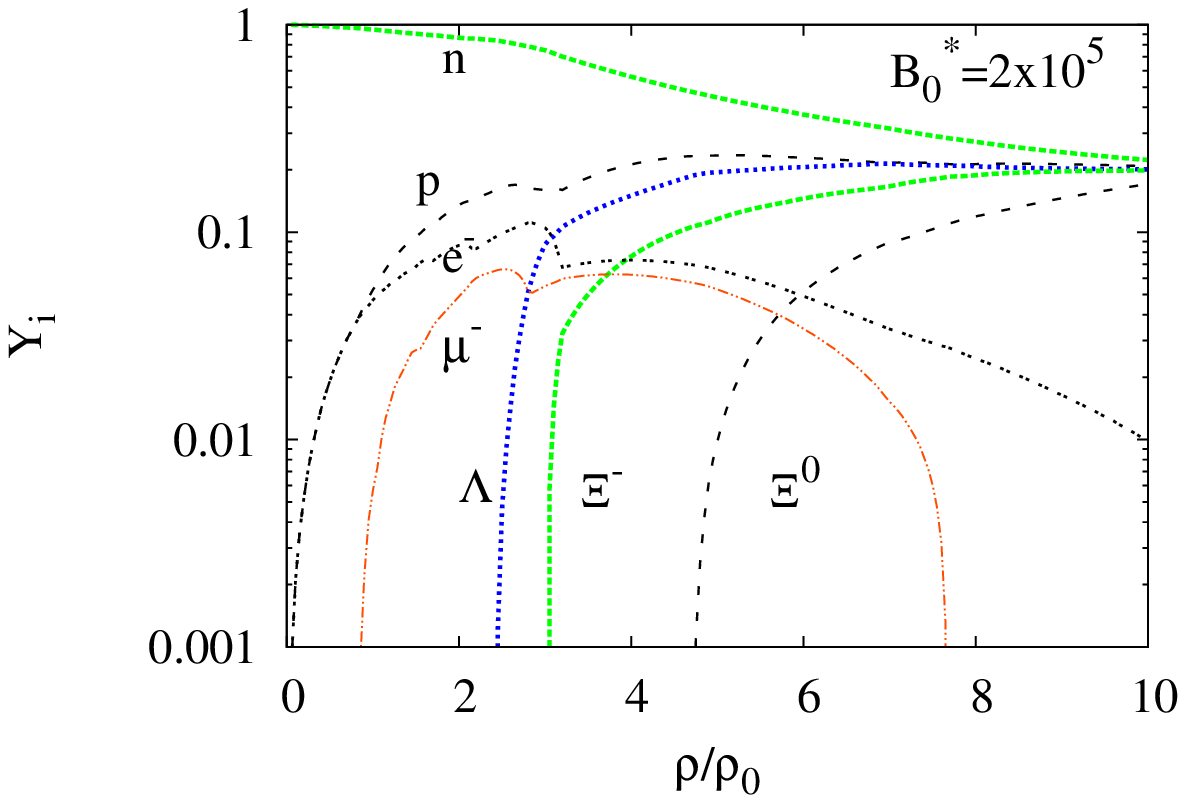}
\includegraphics[width=7.8cm]{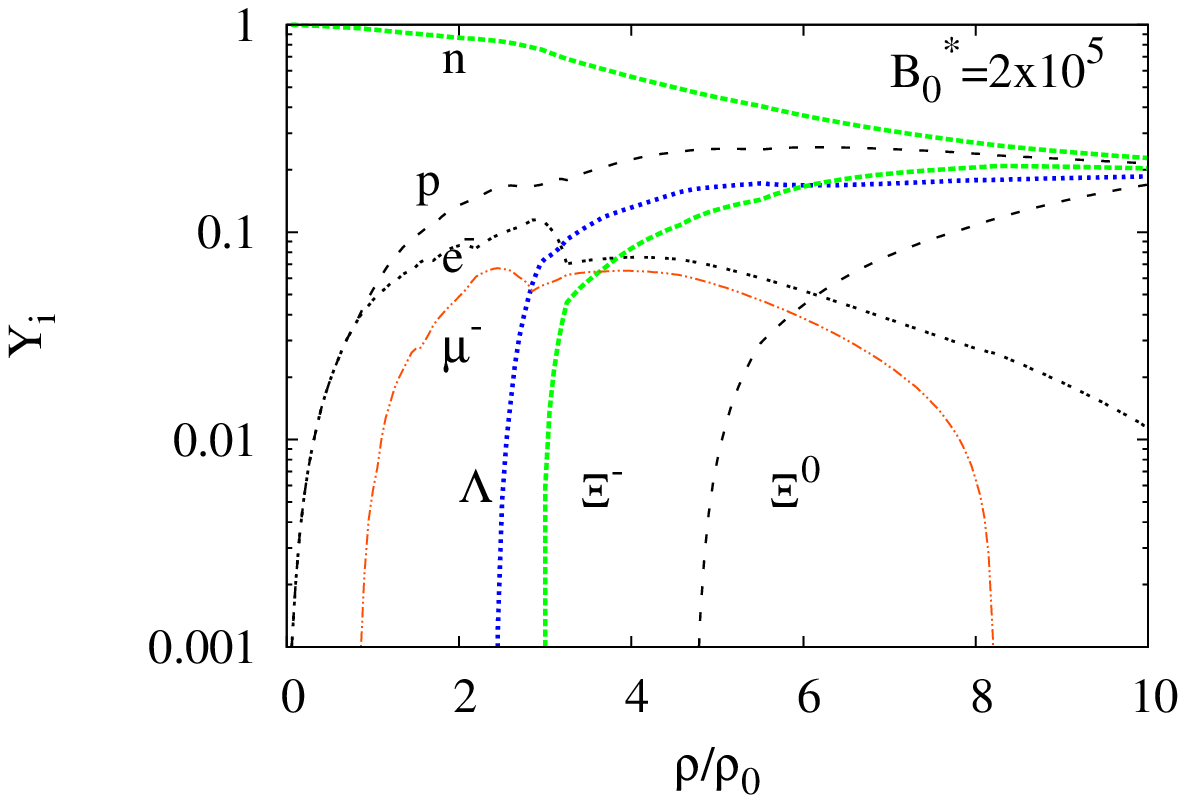}  \\
\includegraphics[width=7.8cm]{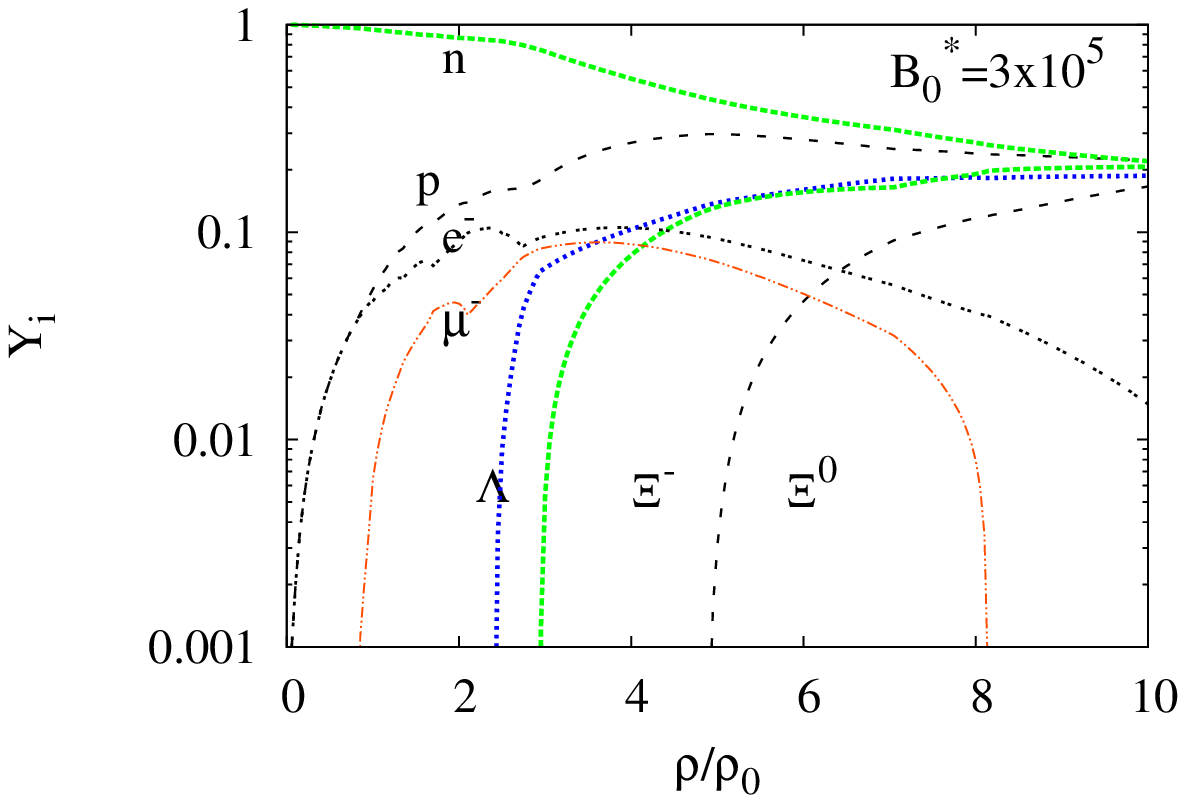}
\includegraphics[width=7.8cm]{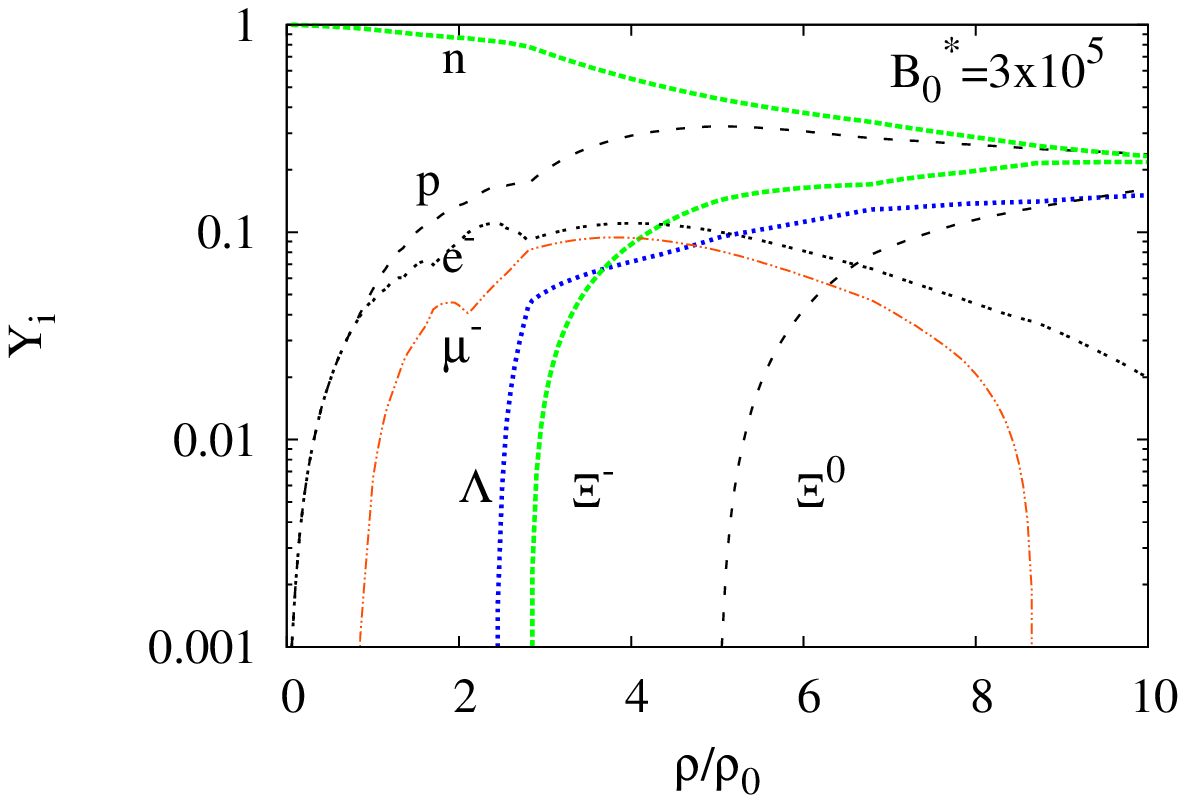}
\caption{(Color online) Same as Fig.1 but for the fast varying magnetic field
($\beta = 0.02$ and $\gamma = 3$). For more direct comparison, all
results for p and $\Xi^-$ are summarized in the right hand side
(RHS) of Fig.3. } \label{fig:popul-f}
\end{figure}
\begin{figure}
\centering
\includegraphics[width=7.8cm]{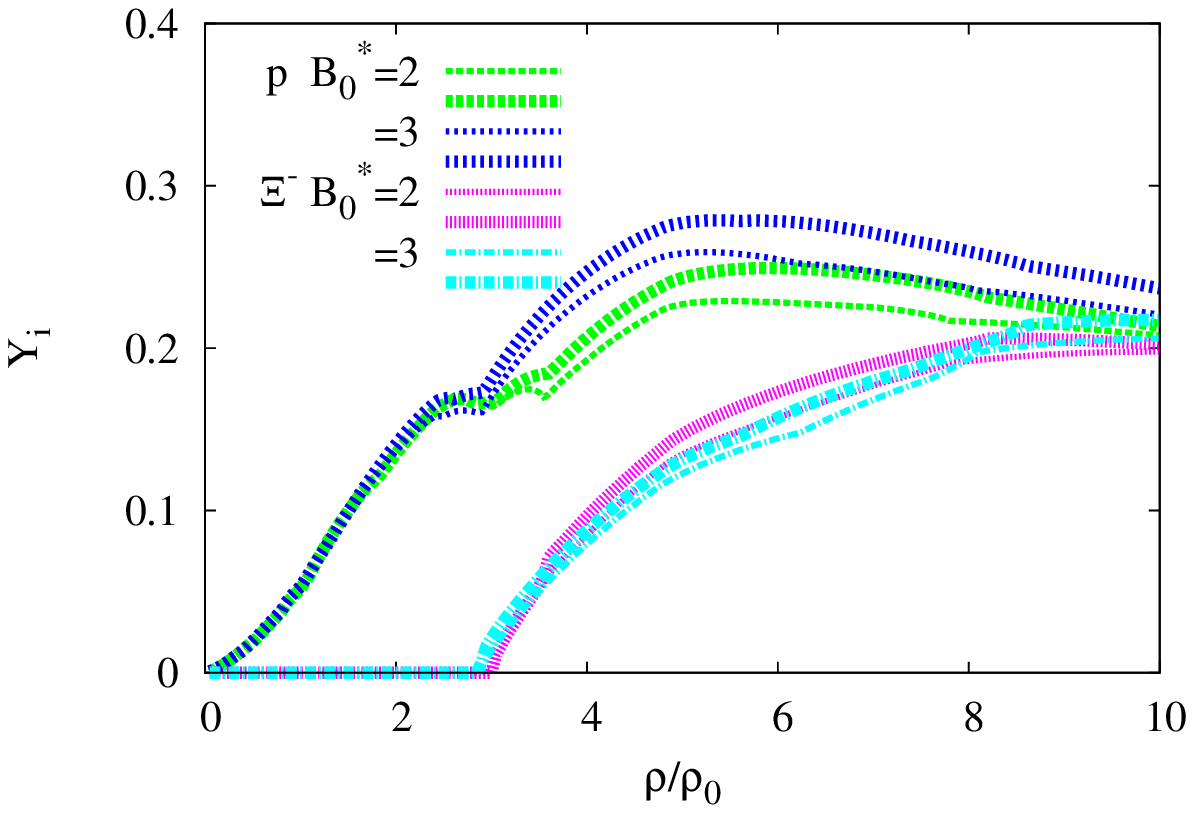}
\includegraphics[width=7.8cm]{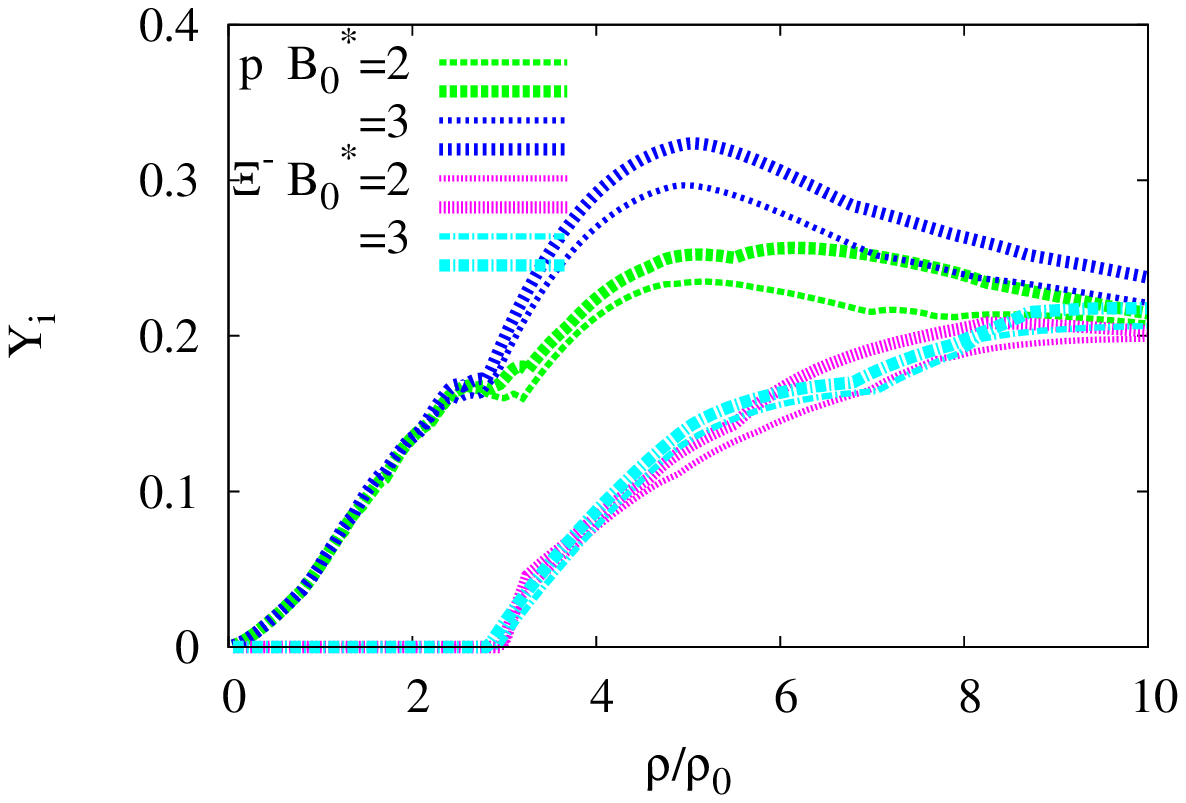}
\caption{(Color online) Populations of p and $\Xi^-$ in a neutron star for both
slow (LHS) and fast (RHS) varying magnetic field. Thick lines
represent results for density-dependent AMM and thin lines are for
constant AMM. $B_0^*$ values are given by a unit of $10^5$.}
\label{fig:popul-pxi}
\end{figure}

Before presenting medium effects by density-dependent AMMs, we
shortly discuss effects of a magnetic field on a neutron star. The
strong magnetic field affects charged particles through the EM
interaction term ($eB$), which leads to the Landau quantization,
and all baryons by the AMM term ($\kappa_b B$).

The quantum numbers for the Landau levels have positive values $\nu =
0, 1, 2 \cdots$, so that the magnetic field increases energies of
charged particles. Consequently, the chemical potentials of
charged particles are increased by the magnetic field.

On the other hand, the AMM term gives rise to the spin splitting,
so that the energy level is divided into two levels: one is higher
level and the other is lower one. Since the chemical potential
means the Fermi surface energy of a particle, the AMM term with
increasing magnetic fields enlarges the chemical potential of a baryon.

If we allow the variations of AMMs in a nuclear medium, AMM values
of relevant baryons are usually swollen. According to our previous
results by the MQMC model \cite{ccts2008}, for example, the AMM
enhancements of a proton, $\Lambda$, and $\Xi$ are about $25 \%$,
$10 \%$ and $5 \%$, respectively, at saturation density.
Therefore, medium effects due to density-dependent AMMs cause
the chemical potentials of relevant baryons to become larger
in addition to the enlargement by the effect of magnetic fields.

In Figs. \ref{fig:popul-s} and \ref{fig:popul-f}, populations of
baryons and leptons for the slow (Fig. 1) and the fast (Fig. 2)
varying magnetic fields are presented for various $B_0^*$ values.
Left panels are results of constant AMMs and right panels are those of
density-dependent AMMs. Populations of protons and electrons are
enhanced with the higher $B^*$ from upper to lower figures. If we
notice the electron population at $\rho/\rho_0$ = 10, the
enhancement is easily discerned. In particular, the population of
electrons is larger than that of protons because the Bohr magneton
$\mu_e$ is about 2000 times larger than the nucleon magneton
$\mu_N$. This effect is fully ascribed to the increased magnetic
fields.

The difference between left and right panels shows medium effects
due to density-dependent AMMs. One can notice the increase of
electron population from left to right panels. The higher magnetic
field is given, the larger medium effect appears.

In order to clearly demonstrate both effects, {\it i.e.} the magnetic
field effects and medium effects due to density-dependent AMMs, we
present both effects in a sheet in Fig. \ref{fig:popul-pxi}. We
showed populations of protons and a $\Xi^-$ for two different
$B^*_0$ fields, and for the constant AMM and the density-dependent
AMM values. Since both effects increase chemical potentials of
charged particles, populations of both particles are clearly
increased.

The magnetic field effect seems to play a major role of increasing
populations compared to the medium effect. But, in the $\rho /
\rho_0 = 6 \sim 8$ region, the medium effect due to
density-dependent AMMs can be competing with the magnetic field
effect. The medium effect is almost same as that by the magnetic
field increased by one unit.

The enhancement of the proton fraction gives rise to the suppression of
other baryons because of baryon number conservation. It means that
there appears the suppression of neutrons and $\Lambda$ hyperons
as shown in Figs. \ref{fig:popul-s} and \ref{fig:popul-f}.

On the other hand, the threshold density for $\Xi^-$ is pushed to
the higher density with the stronger magnetic field as shown in
Fig. \ref{fig:popul-pxi}. However, the abundance of $\Xi^-$ is not changed so much in
comparison with $\Lambda$ as shown in Figs. \ref{fig:popul-s} and
\ref{fig:popul-f}. Since the $\Xi^-$ hyperon is a charged
particle, the population is increased by the magnetic field, while
the baryon number conservation and the charge neutrality lead to
suppress the population. Therefore, the behavior of $\Xi^-$
population is balanced by the effects of the magnetic field and
the conditions of a neutron star.

The difference between the slow (Fig. \ref{fig:popul-s}) and fast
(Fig. \ref{fig:popul-f}) varying magnetic fields is the slope of
magnetic field in the region of middle densities. Therefore, this
difference just corresponds to the increase of the magnetic field
strength $B^*_0$ at the same density. However, the effect due to
the difference is not remarkable because the exponential term is
small by comparing with the effect of the $B_0^*$ term,
irrespective of $\gamma$ and $\beta$ values used here.

\subsection{Medium Effects on the EoS, Mass and Radius}
\begin{figure}
\centering
\includegraphics[width=7.8cm]{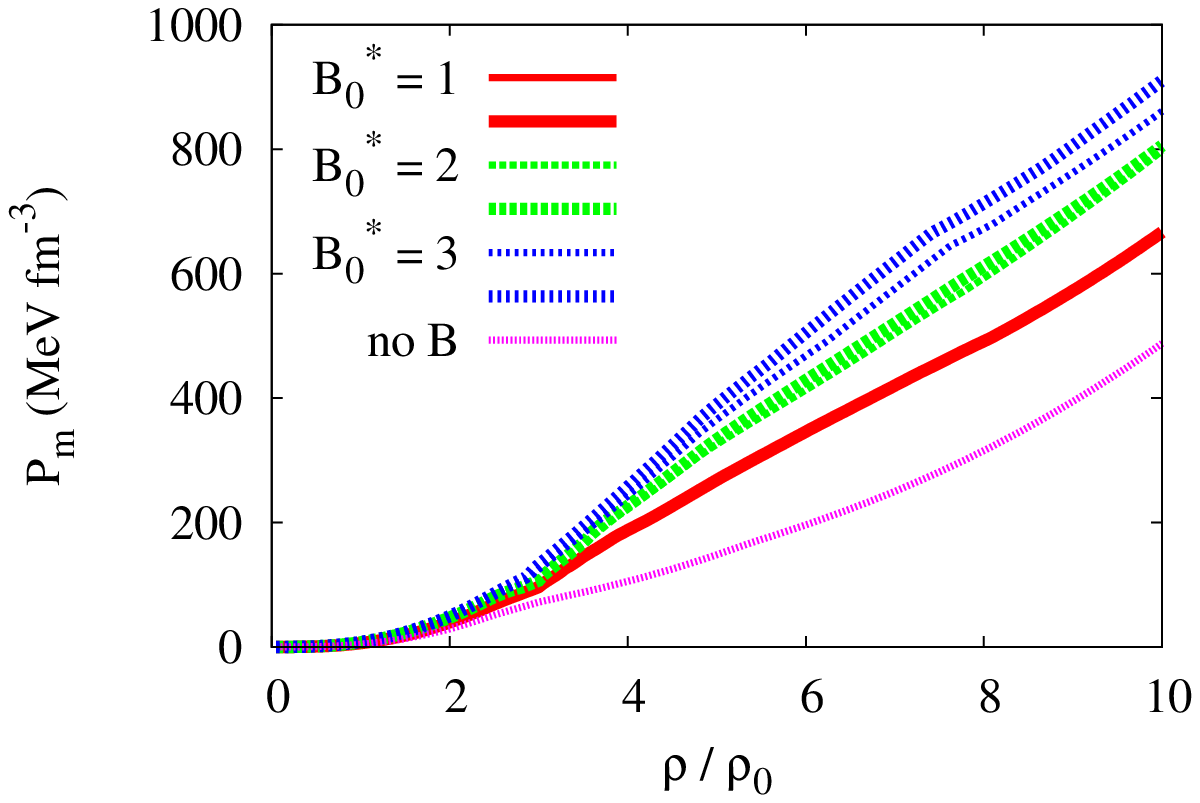}
\includegraphics[width=7.8cm]{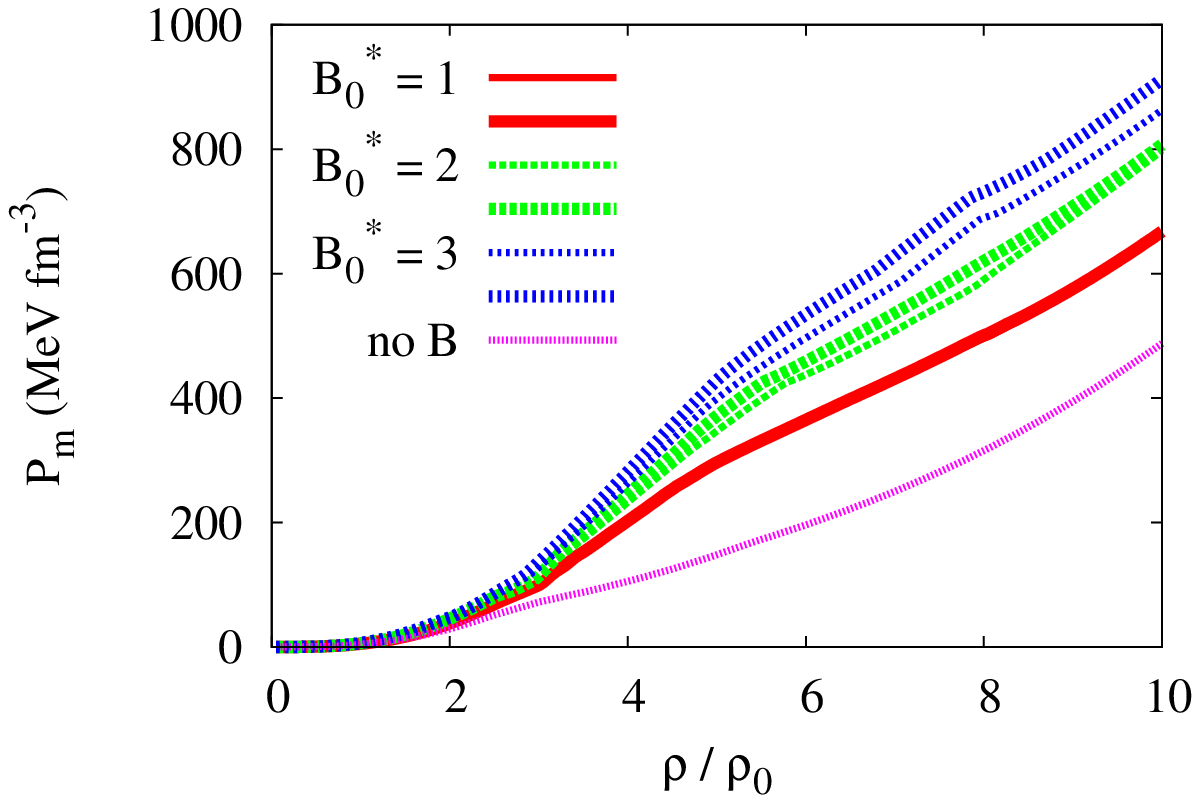}
\caption{(Color online) Equation of state (slow in LHS and fast in RHS). Thick
lines represent results for density-dependent AMM and thin lines are for
constant AMM. $B_0^*$ values are given by a unit of $10^5$.} \label{fig:eos}
\end{figure}
\begin{figure}
\centering
\includegraphics[width=7.8cm]{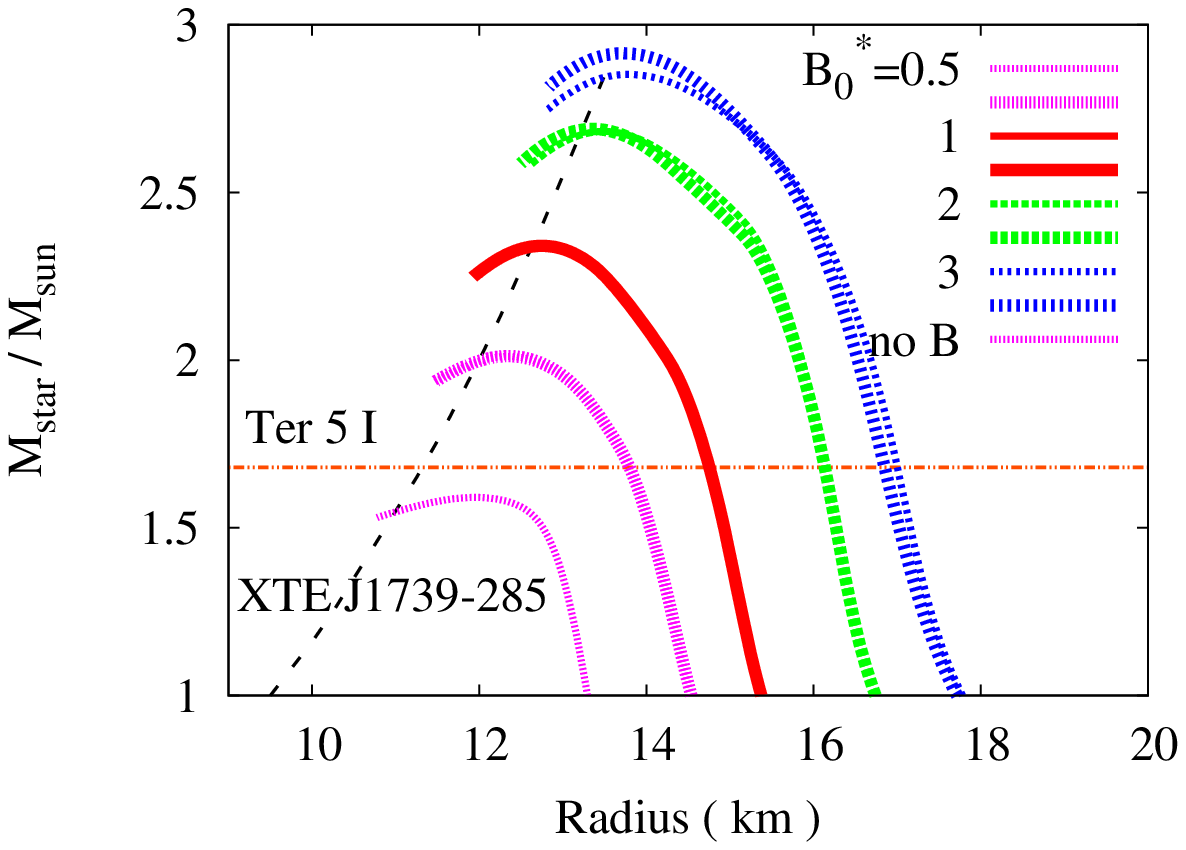}
\includegraphics[width=7.8cm]{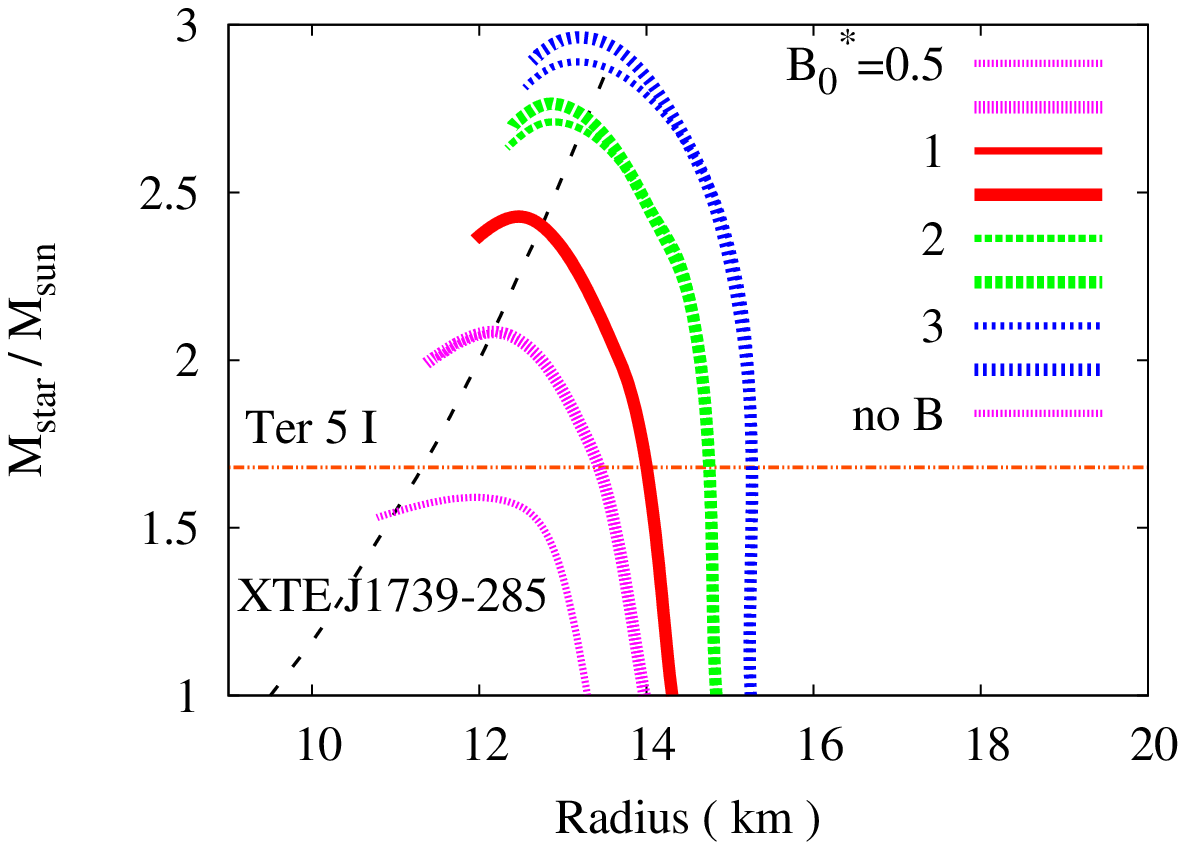}
\caption{(Color online) Mass-radius relations.
Thick lines
represent results for density-dependent AMM and thin lines are for constant
AMM. LHS is for the slow varying magnetic field and RHS is for the
fast one. $B_0^*$ values are given by a unit of $10^5$.} \label{fig:mass}
\end{figure}

Magnetic fields and density-dependent AMMs also affect the EoS and
the maximum mass of a neutron star. As shown in Fig.
\ref{fig:eos}, the EoS in dense matter becomes stiffer with the
increase of chemical potentials and the suppression of hyperons by
the magnetic field. As a result, maximum masses of neutron stars
are increased. The pressure due to matter fields also strongly
depends on the strength of magnetic fields, but weakly depends on
the density-dependent AMMs as shown in Fig. \ref{fig:eos}. For the
fast magnetic field, slopes of EoS between $\rho / \rho_0 = 3$ and
$5$ are rapidly changed because magnetic fields cause the EoS to be
fast stiffer.

But the effects of density-dependent AMMs are smaller than those of the
magnetic field strength similar to the case of the populations. In
a relatively small magnetic field ($B_0^* = 1 \times 10^5$ G), the
density-dependence of AMMs rarely affects the EoS. However, in the
strong magnetic fields ($B_0^* \geq 2 \times 10^5$), the
contribution of the density-dependent AMMs in nuclear medium
appears explicitly. For example, the increase of the pressure
$P_m$ is about 37 MeV fm$^{-3}$ at $\rho = 6 \rho_0$ for the fast
case in $B_0^* = 3 \times 10^5$.

Mass-radius relations of neutron star obtained from TOV equations
are shown in Fig. \ref{fig:mass}. Masses of neutron stars,
$M_{star}$, which are obtained from total energy density and total
pressure ($\varepsilon_{tot}$, $P_{tot}$), depend very strongly on
the strength of magnetic fields. But the contribution of
density-dependent AMMs is indiscernible, which is about $0.1
M_\odot$, maximally, even in the largest magnetic fields.
Since there are no direct data for mass-radius relation of magnetars,
we compare our results with the observed neutron stars in next section.

\subsection{Comparison with observations}
Neutron stars and heavy ion collisions may provide valuable
constraints for the nuclear EoS \cite{klahn2006}. Recent data
reported higher masses and larger radii for neutron stars. For
instance, $M = 2.0 \pm 0.1 M_\odot$ for 4U 1636-536 was reported in Ref. \citep{lava2008} and
authors in Ref. \cite{steiner} recently investigated seven neutron stars,
six binaries and a isolated neutron star (RX J1865-3754), showing $M = 1.9-2.3 M_\odot$
and $R = 11 - 13$ km.
Pulsar I of the globular cluster Terzan 5 (Ter 5 I) shows a lower
mass limit $M \ge 1.68 M_\odot$ at 95 \% confidence level
\cite{ransom}. Another constraint deduced independently of given
models is obtained from XTE J1739-285 \cite{kaaret}, which
presents a constrained curve for the ratio between mass and
radius.

Thus we compare our results with Ter 5 I and XTE J1739-285 in Fig.
\ref{fig:mass}. In the hyperonic star without magnetic fields ('no
B' in Fig. \ref{fig:mass}), the maximum mass is about 1.59
$M_\odot$ which does not satisfy the mass limit ($1.68 M_\odot$)
by Ter 5 I. In addition, the constraint by XTE J1739-285 runs
though an unstable region. When magnetic fields are introduced,
the LHS in Fig. \ref{fig:mass} for the slow varying field shows
that the line by XTE J1739-285 also goes through the unstable
region. However, results for the fast varying magnetic field can
satisfy the constraint of XTE J1739-285 and explain masses of
neutron stars as $2-3$ $M_\odot$ with magnetic fields for hyperonic
stars.

In order to detail density-dependent AMMs effects, in Tab.
\ref{tab:result}, the central density ($\rho_c$), maximum masses,
and magnetic fields at central density ($B^*_c$) for the fast
varying magnetic field are tabulated for both constant and
density-dependent AMMs cases. The effects of density-dependent AMMs are
negligible in small magnetic fields. But as magnetic fields
increase, the effect also increases and then the maximum mass is
increased by about 0.07 $M_\odot$ for $B_0^* = 3 \times 10^5$ in
the fast varying magnetic fields.

\begin{table}
\begin{center}
\begin{tabular}{ c|cccc } \hline
      ~~ &~~~~~$B_0^*$~~~~~&  $\rho_c$ & $M_{star} / M_\odot$ &    $B_c^*$         \\ \hline
      &~~~~$5 \times 10^4$~~~~&~~~~~~6.05~~~~~~&~~~~~~2.08~~~~~~&~~~~$4.94 \times 10^4$~~~~\\
Constant & $1 \times 10^5$ & 6.05      & 2.43          & $9.88 \times 10^4$ \\
~~~AMM~~~& $2 \times 10^5$ & 5.55      & 2.71          & $1.93 \times 10^5$ \\
         & $3 \times 10^5$ & 4.90      & 2.88          & $2.71 \times 10^5$ \\  \hline
         & $5 \times 10^4$ & 6.05       & 2.08          & $4.94 \times 10^4$ \\
density-dependent & $1 \times 10^5$ & 6.05       & 2.42          & $9.88 \times 10^4$ \\
   AMM   & $2 \times 10^5$ & 5.75       & 2.76          & $1.95 \times 10^5$ \\
         & $3 \times 10^5$ & 5.00       & 2.96          & $2.75 \times 10^5$ \\ \hline
\end{tabular}
\end{center}
\caption{The central density ($\rho_c)$, maximum masses ($M /
M_\odot$) and central magnetic field ($B_c^*$) for various
$B_0^*$ in both constant and changing AMMs.
The results are obtained from fast varying magnetic fields.}
\label{tab:result}
\end{table}

Finally, one can derive the limit of magnetic fields in the
interior of a neutron star and the limit of density-dependent AMMs
in medium. The allowed strength of magnetic fields is usually
constrained by the scalar virial theorem \cite{su2001,BWN}. It is
given by the following approximate relation, $B \sim 2 \times 10^8
(M / M_\odot) (R_\odot / R)^2$ G for the non-rotating star. For
the star with $R \approx 10$ km and $M \sim M_\odot$, we obtain $B
\sim 10^{18}$ G from the above relation.

In the calculation of model independent method for the maximum
mass of neutron star, the limit of maximum mass is about $M = 3
\sim 5 M_\odot$ \cite{BWN}. Furthermore the observations show
that there is no any neutron star in large mass region, which
exceeds 3 $M_\odot$. In this results, for fast case in $B^*_0 = 3
\times 10^5$ G, the maximum mass of the star is $2.96(2.89)
M_\odot$ for density-dependent (constant) AMM and the central magnetic
fields is about $B = 2.75(2.71) \times 10^5 B_e^c$ G. We can thus
conclude that the upper limit of magnetic fields might be $B
\approx 3 \times 10^5 B_e^c$ G in neutron star with hyperons in
this work, although detailed numbers depend on the model and
parameters.

According to the model dependence of the AMM in other calculations
\cite{meisner,cheon92,frank96,lu99,yak03,smith04,horikawa05}, the largest
enhancement is by about $40 \%$ for nucleons at saturation density
\cite{cheon92}, but the other models show the enhancement of about
$10 \sim 25 \%$. Thus the enhancement of $25 \%$ in this work,
corresponds to the maximum enhancement except Ref. \cite{cheon92}.
If we employ much larger enhancement for the AMM like the value in
Ref. \cite{cheon92}, the contribution of density-dependent AMM in
medium may become larger. However, all populations, EoS, and maximum
mass should depend on the strength of magnetic field very strongly,
so that the contribution due to varying the AMM is still remained as a
subsidiary role. Thus the effect of density-dependent AMM might be
maximally around $0.1 M_\odot$.

\section{Summary}
We investigate the effect of the density-dependent AMM of baryons
in neutron star under strong magnetic fields by using the QHD
model, which includes baryon octet and leptons. By exploiting the
density-dependent AMM values of baryons obtained from the MQMC
model, we calculate the populations of particles, EoS, and the
mass-radius relations for the slow and the fast varying magnetic
fields. The strength of magnetic field is expressed as EM
interaction of all charged particles and their AMM of baryon
octet.

In the populations of particles, all charged particles experience
Landau quantization and its effect depends severely on the
strength of magnetic fields. The increase of the magnetic fields
enhances the chemical potentials of all charged particles. In
particular, since a proton is the lightest particle among baryons,
the fraction of protons is enlarged by the magnetic field. As a
result, it gives to rise the suppression of hyperons to satisfy
the conservation of baryon number.
The EoS becomes stiffer and then maximum mass of neutrons star also becomes larger.

The mass-radius relations of neutron star obtained from magnetic fields
are compared with observational data. The mass-radius relation by fast varying magnetic fields
satisfy the constraint by XTE J1739-285. The effect of density dependent AMM
appears in very high magnetic fields, causing the increase of maximum mass of the star
with about 0.1 $M_\odot$ in $B_0^* = 3 \times 10^5$.

We assume the constant magnetic field along z-axis for
non-rotating star. However, the real neutron star under the strong
magnetic field rotates very rapidly and the magnetic fields may be
taken place by the rotation of matter fields \cite{yasu2009}. Thus
the calculation should be self-consistent with each other, that
is, the matter fields in rotating star create magnetic field and
the produced magnetic field affects matter fields. This
self-consistent approach for the magnetic field will be our next
work.

\section*{ACKNOWLEDGMENTS}
This work was supported by the Korea Science and Engineering
Foundation (KOSEF) Grant funded by the Korean Government
(R01-2007-000-20569) and the Soongsil University Research Fund.
Work of CYR was supported by the Korea Research Foundation Grant
funded by the Korean Government (MOEHRD) (KRF-2008-214-C00015).

\end{document}